\documentstyle[11pt,epsfig]{article}
\date{\empty}
\renewcommand{\abstract}[1]{{ \footnotesize \noindent {\bf Abstract} #1 \\}}
\renewcommand{\author}[1]{\subsubsection*{\it#1}}
\newcommand{\address}[1]{\subsubsection*{\it#1}}

\begin{document}

\title{Interstellar dust evolution in galaxies of different morphological types 
\label{calura}}

\maketitle

\author{Francesco Calura$^1$\\
Antonio Pipino$^{2,3}$ \\
Francesca Matteucci$^{1,2}$\\
}
\address{$1$ INAF - Osservatorio Astronomico di Trieste, via G. B. Tiepolo 11, 34143 Trieste, Italy \\
$2$Dipartimento di Astronomia - Universit\'a di Trieste, Via G. B. Tiepolo 11, 34143 Trieste, Italy \\
$3$ Astrophysics, University of Oxford, Denys Wilkinson Building, Keble Road, Oxford OX1 3RH, U.K.
}

\abstract{We study interstellar dust evolution in various environments 
 by means of chemical evolution models for galaxies of different morphological types.  
We start from the formalism developed by Dwek\cite{DWE98}  to study dust evolution in the solar neighbourhood and 
extend it to ellipticals and dwarf irregular galaxies, showing how 
the evolution of the dust production rates and of the dust fractions depend on the galactic star formation history.  
The observed dust fractions observed in the solar neighbourhood can be reproduced by assuming that 
dust destruction depends the condensation temperatures $T_{c}$ of the elements. 
In elliptical galaxies, type Ia SNe are the major dust factories in the last 10 Gyr. 
With our models, we successfully reproduce the dust masses observed in local ellipticals ($\sim 10^{6}M_{\odot}$)
by means of recent FIR and SCUBA observations. 
We show that dust is helpful in solving the iron discrepancy observed in the 
hot gaseous halos surrounding local ellipticals. In dwarf irregulars, we show how 
a precise determination of the dust depletion pattern could be useful to put solid constraints on the dust condensation efficiencies.  
Our results will be helpful to study the spectral properties of dust grains in local and distant galaxies.}

\bigskip
\bigskip
\section{Introduction}
The presence of interstellar dust 
affects strongly the spectral properies of galaxies, in particular in the ultra-violet (UV) band 
and in the far infrared (FIR). In the UV, the stellar light is absorbed and scattered (dust extinction), 
whereas in the FIR, dust grains thermally re-emit most of the energy  absorbed in the UV. \\
However, the presence of interstellar dust can affect also the chemical composition of the interstellar medium (ISM). 
Some chemical elements, called refractory (such as C, O, Fe), in the gas phase undergo 
dust depletion and a fraction of their total abundance 
is incorporated into dust  grains. For these elements, the abundance measured in stellar 
surfaces is considerably higher than the gas phase 
abundances (\cite{KIM03}, \cite{ZUB04}, \cite{DRA03}).\\  
In this paper, we aim at studing the effects of dust grains on the chemical composition of the ISM by means 
 of chemical evolution models for galaxies of different morphological types. 
We start from the formalism developed by \cite{DWE98}, and extend it to environments different than the solar neighbourhod, 
i.e. to elliptical galaxies and dwarf irregular galaxies. In particular, we focus on the effects that different star formation histories 
have on the evolution of the dust content of galaxies of different morphological types.\\
This contribution is organized as follows. 
In section 2 we present the chemical evolution models. 
In section 3, we discuss our results for the S.N., for elliptical galaxies and for dwarf 
irregular galaxies. Finally, in Section 4 we present our summary.   

\section{Chemical evolution of the interstellar dust}

\subsection{Chemical evolution models for galaxies of different morphological types} 
Chemical evolution models are used to to follow the evolution of the abundances of several chemical
species and of the dust content of spirals, elliptical and irregular galaxies. \\
In our picture, elliptical galaxies form as a result of the rapid collapse of a homogeneous sphere of
primordial gas where intense star formation (SF) is taking place at the same time as the collapse proceeds. 
SF is assumed to halt as the energy of the ISM, heated by stellar winds and supernova (SN) explosions,
exceeds the binding energy of the gas. At this time a galactic wind occurs, sweeping away almost all of  
the residual gas. After the SF has stopped, the galactic wind is maintained by type Ia SNe, and its duration
depends on the balance between this heating source and the gas cooling (we refer the reader to \cite{PIP02}, \cite{PIP05}).\\ 
In our scheme, large galaxies form the bulk of their stars and develop a galactic winds on shorter timescales than small galaxies, 
according to the ``inverse wind'' scenario \cite{MAT94}. 
Here, the models La1 and Ha1 of \cite{PIP05} are  used for an elliptical galaxy of  
luminous mass $M_{\rm lum}=10^{11}M_{\odot}$ and $M_{\rm lum}=10^{12}M_{\odot}$, respectively. \\
For spiral galaxies, the adopted model is calibrated in order to reproduce a large set of observational 
constraints for the Milky Way galaxy \cite{CHI01}. 
The Galactic disc is approximated by several independent rings, 
2 kpc wide, without exchange of matter between them. In our picture, 
spiral galaxies are assumed to form as a result of two main infall episodes.  
During the first episode, the halo and the thick disc are formed.
During the second episode, a slower infall
of external gas forms the thin disc with the gas accumulating faster in the inner than in the outer
region ("inside-out" scenario, \cite{MAT89}. The process of disc formation is much longer than the
halo 
and bulge formation, with time scales varying from $\sim2$ Gyr in the inner disc to $\sim7$ Gyr in the solar region
and up to $20$ Gyr in the outer disc (see table 1). In this paper, we are interested in the study of 
dust evolution in the solar neighbourhood (S.N.). For this purpose, we focus on a ring located at 8 kpc from the 
Galactic centre, 2 kpc wide. \\ 
Finally, irregular galaxies are assumed to assemble from infall of protogalactic small clouds 
of primordial chemical composition, until masses in the range $\sim 10^{9}M_{\odot}$ are accumulated, 
and to produce stars at a lower rate than spirals. \\
In figure 1, we show the star formation histories of the galactic models used in this work. \\

\begin{figure*}
\centering
\vspace{0.001cm}
\epsfig{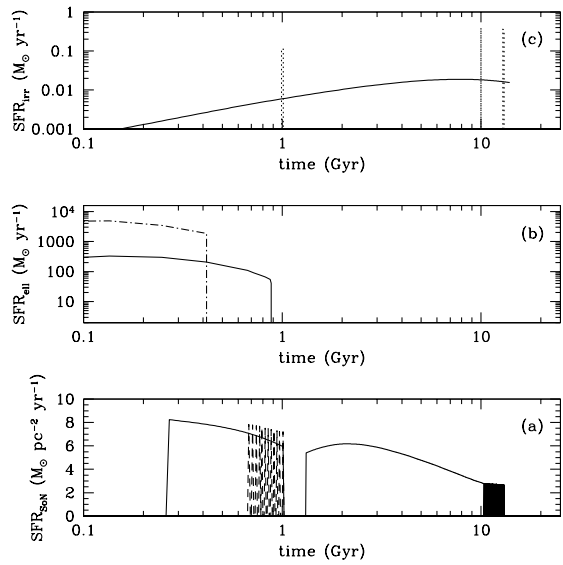}
\caption[]{Predicted SFRs as a function of time for different chemical evolution models. Panel (a): 
S.N. (solid line) and a model for the outer regions of the Milky Way disc (dashed line); 
panel (b): two different elliptical galaxy models, i.e. model La1 (solid line) and Ha1 (dash-dotted line) of \cite{PIP05} ; 
panel (c): two different irregulars, one with continuous SF (solid line) and the 
other with 3 starbursts (dotted line). 
}
\label{fig1}
\end{figure*}

The chemical enrichment of the ISM is due to three contributors: low and intermediate mass stars 
(i.e. with masses $0.8 M_{\odot}\le m \le  8 M_{\odot}$); type Ia SNe, assumed to originate from exploding white dwarfs in binary systems,
with  
total masses $3 M_{\odot} \le m_{bin} \le 16 M_{\odot}$ \cite{GRE83},\cite{MAT86}, 
and single massive stars with initial masses $m>8 - 100 M_{\odot}$ explode as core collapse SNe.  \\
The nucleosynthesis prescriptions are common to all models. 
For massive stars and type Ia SNe, we adopt the empirical yields suggested by  \cite{FRA04},   
which are substantially based on the \cite{WOO95} and \cite{IWA99} yields, respectively, 
and are tuned to reproduce at best the abundances in the S.N.  
For low and intermediate mass stars (LIMS), we adopt the prescriptions by \cite{VAN97}.  \\
We adopt a Salpeter \cite{SAL55} IMF for ellipticals and irregulars and a Scalo \cite{SCA86} IMF for the spiral model.\\
It is worth to stress that the set of chemical evolution models presented in this paper allows us to reproduce 
the main observational features of local galaxies 
\cite{CAL06a}, and to account 
for the local metal budget \cite{CAL04}, 
as well as  for the cosmic history of metal production \cite{CAL06b}
and star formation (\cite{CAL03}; \cite{CAL04a}).

\subsection{How to model dust production, dust destruction and dust accretion} 
The chemical evolution of an element $i$ in the dust is computed by using 
the formalism developed by \cite{DWE98}. 
Let $X_{dust, i} (t)$ be the abundance by mass of the element \emph{i} in the dust and 
$G(t)$ the ISM fraction at the time $t$, the quantity 
$G_{dust,i}(t)= X_{dust,i} \cdot G(t)$
represents  the normalized mass density
of the element \emph{i} at the time \emph{t} in the dust. The time evolution 
of $G_{dust,i}(t)$ is calculated by means of the following equation:

\begin{eqnarray}
 & & {d G_{dust,i} (t) \over d t}  =  -\psi(t)X_{dust,i}(t)\nonumber\\
& & + \int_{M_{L}}^{M_{B_m}}\psi(t-\tau_m) \delta^{SW}_{i}
Q_{mi}(t-\tau_m)\phi(m)dm \nonumber\\ 
& & + A\int_{M_{B_m}}^{M_{B_M}}
\phi(m)\nonumber \\
& & \cdot[\int_{\mu_{min}}
^{0.5}f(\mu)\psi(t-\tau_{m2}) \delta^{Ia}_{i}
Q_{mi}(t-\tau_{m2})d\mu]dm\nonumber \\ 
& & + (1-A)\int_{M_{B_m}}^
{8 M_{\odot}}\psi(t-\tau_{m})  \delta^{SW}_{i} Q_{mi}(t-\tau_m)\phi(m)dm\nonumber \\
& & + (1-A)\int_{8 M_{\odot}}^
{M_{B_M}}\psi(t-\tau_{m})  \delta^{II}_{i} Q_{mi}(t-\tau_m)\phi(m)dm\nonumber \\
& & + \int_{M_{B_M}}^{M_U}\psi(t-\tau_m)  \delta^{II}_{i} Q_{mi}(t-\tau_m) 
\phi(m)dm \nonumber\\ 
& & - \frac{G_{dust,i}}{\tau_{destr}} + \frac{G_{dust,i}}{\tau_{accr}} -({d G_{dust,i} (t) \over d t})_{\rm out}  	
\label{eq_dust}
\end{eqnarray}

(for a detailed explanatio of Eq. 1, see \cite{CAL08}. 
$\psi(t)$ is the star formation rate.  The second, third, fourth, fifth and sixth terms of Eq. 1 represent the contributions 
by stellar sources of various types. These terms contain the quantities 
$\delta^{SW}_{i}$, $\delta^{Ia}_{i}$ and  $\delta^{II}_{i}$, which  are the condensation 
efficiencies of the element $i$ in stellar winds, type Ia and type II SNe and regulate the stellar production of interstellar dust.  
These quantities represent the 
fractions of the element $i$ which is condensed into dust and restored into the ISM by all the stellar types. 
For the condensation efficiencies, we use the same prescriptions as described in \cite{DWE98}. 
In general, we assume that stars can produce two different types of dust grains:   
silicate dust, composed by O, Mg, Si, S, Ca, Fe, and carbon dust, composed by C. 
The seventh and eighth terms of eq.~\ref{eq_dust} represent  the dust destruction and accretion rates, respectively. 
These terms depend on the quantities  $\tau_{destr}$ and $\tau_{accr}$, which represent the typical 
timescales for destruction and accretion, 
respectively. \\
Dust destruction is assumed to occurr in SN shocks. 
Following the suggestions by \cite{MCK89} and \cite{DWE98}, for a given element $i$ 
the destruction timescale $\tau_{destr}$ can be expressed as: 
\begin{equation}
\tau_{destr, i}=(\epsilon M_{SNR})^{-1} \cdot \frac{\sigma_{gas}}{R_{SN}}
\end{equation} 
$R_{SN}$ is the total SNe rate, 
including the contributions by both type Ia and type II SNe. 
$M_{SNR}$ is the mass of the interstellar gas swept up by the SN remnant. 
For this quantity, \cite{MCK89} suggests 
a typical value of $ M_{SNR} \sim 6800 M_{\odot}$, which is in agreement 
with the results from numerical studies of SN evolution  \cite{THO98}. 
As suggested by \cite{MCK89}, typical 
values for the destruction efficiency $\epsilon$  in a three-phase medium 
as the present-day local ISM are around 0.2, hence we assume:  
\begin{equation}
\epsilon M_{SNR} = 0.2 \times 6800 M_{\odot} = 1360 M_{\odot}
\, .\\
\end{equation}
Dust accretion is assumed to occur in dense molecular clouds, 
where volatile elements can condensate onto pre-existing grain cores, originating a volatile 
part called mantle (\cite{DWE98}, \cite{INO03}). 
For a given element $i$,  
the accretion timescale $\tau_{accr}$ can be expressed as: 

\begin{equation}
\tau_{accr}=\tau_{0,i}/(1 - f_i) 
\label{accr_t}
\end{equation} 

where 
\begin{equation}
f_i=\frac{G_{dust,i}}{G_{i}}
\end{equation}

According to eq.~\ref{accr_t}, the accretion timescale is an increasing function of the dust mass. 
For the timescale $\tau_{0,i}$, typical values span from $\sim 5 \times 10^{7}$ yr, of the order of 
the lifetime of a typical molecular cloud \cite{DWE98}, up to $\sim 2 \times 10^{8}$ yr  \cite{DWE98}. 
In this paper, we assume a timescale $\tau_{0,i} = 5 \times 10^{7} yr$.\\ 
In elliptical galaxies, we assume that dust accretion occurs only during the starburst epoch, when large 
amounts of cold gas and 
molecular H are available. \\
The observed molecular H content in dwarf irregular galaxies is very small, with molecular-to-atomic gas fractions 
of $\sim 10 \%$ or lower (\cite{LIS98}, \cite{CLA96}). 
Motivated by these observational results, we assume that no accretion can occur in irregular galaxies. 
Finally, the last term of eq.~\ref{eq_dust}  accounts for possible ejection of dust 
into the inter galactic medium (IGM) by means of galactic winds. 
This term is absent in the equation for the S.N. model, 
but is taken into account in the elliptical and irregular models. \\

\section{Results}

\subsection{Dust evolution in the Solar Neighbourhood}

\begin{figure*}
\centering
\vspace{0.001cm}
\epsfig{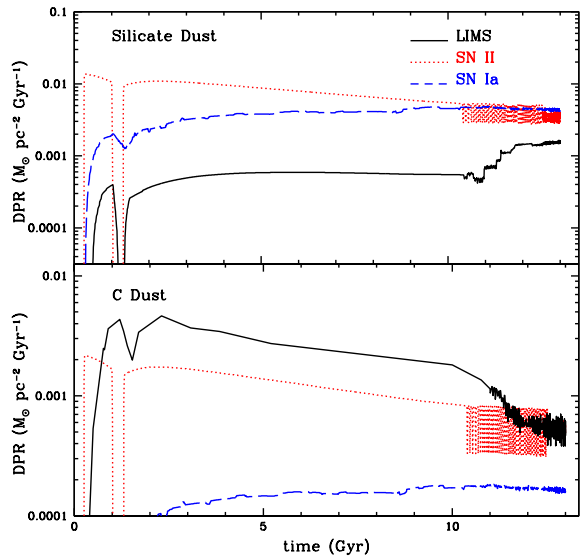}
\caption[]{Predicted dust production rates from various sources 
for a chemical evolution model of the S.N.  
In the lower (upper) panel, we show the results for the carbon (silicate) dust. 
Solid lines: contribution by low and intermediate mass stars (LIMS). Dotted lines: contribution 
by type II SNe. Dashed lines: contribution by type Ia SNe. }
\label{prates}
\end{figure*}

In Figure~\ref{prates}, we show the predicted evolution of the carbon (lower panel) and silicate (upper panel)  
dust production rates calculated by means of the chemical evolution model for the (S. N.). 
Concerning the C dust, its production is dominated by LIMS throughout most of the cosmic time. 
At the present day ($T_{0}\sim 13$ Gyr), a significant contribution is also due to type II SNe, with type Ia SNe 
playing  a negligible role. The frequent discontinuities in the lines are due to the effect of the star 
formation threshold.\\
The production of silicate dust is dominated by type II SNe throughout most of the time. At late times, the contributions by type Ia SNe 
and  type II SNe are comparable. On the other hand, LIMS are negligible contributors to the Si dust production rate. 

\begin{figure*}
\centering
\vspace{0.001cm}
\epsfig{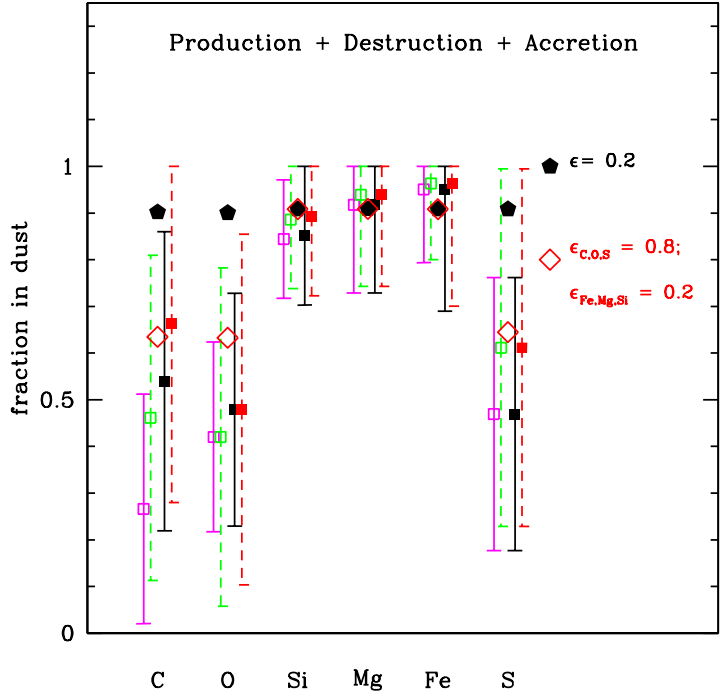}
\caption[]{Fractions in dust for various elements. 
The solid pentagons are the predicted present-day fractions calculated by assuming a 
dust destruction efficiency of $\epsilon=0.2$ for all elements.
The open diamonds are the present-day fractions calculated assuming a 
dust destruction efficiency of  $\epsilon_{C,O,S}=0.8$ for C, O and S and  whereas for 
$\epsilon_{Fe,Mg,Si}=0.2$ for Fe, Si and Mg. 
The solid and open squares are the fractions observed by \cite{KIM03} in the Local Interstellar 
Cloud using the set of cosmic abundances specified in their Tables 2 and 3, respectively. }
\label{fract_epsi}
\end{figure*}

In Fig.~\ref{fract_epsi} we show the predicted present-day fractions in dust, i.e. the ratios between the amount of a given element locked into dust 
and its total ISM abundance, for the 
elements studied in this work, compared to the values observed by \cite{KIM03} in the Local Interstellar Cloud. 
It is worth to note that the Local Interstellar Cloud consists of a warm medium, characterized by a temperature of $\sim 6000 K$, 
whereas our models provide the depletion of the cold gas, with temperatures lower than $\sim 100 K$. 
In this work, the dust fractions observed by \cite{KIM03} are taken as reference. 
From the logarithmic depletions $\delta$ plotted in Figure 2 of \cite{KIM03}, 
it is possible to derive the dust fractions of the cold gas through the formula 
\begin{equation} 
f=1-10^{\delta}
\end{equation} 
For C, O and S, the logarithmic depletions of the cold medium are very similar to the ones of the Local Interstellar Cloud. 
For Si, Mg and Fe, the dust fractions of the cold medium are higher than the ones of the warm medium and are within the error bars 
plotted in Figure~\ref{fract_epsi}. 
As shown by \cite{CAL08}, the dust fractions are nearly independent from 
the choice of the dust condensation efficiencies and with the assumption of a constant 
destruction efficiency $\epsilon$ it is not possible to obtain a satisfactory fit to the observed dust fractions for all elements, in particular for 
C, O, and S. The observed dust fractions can be reproduced by assuming that the dust 
destruction efficiency depends on the physical properties of the chemical element, in particular on the 
condensation temperature $T_{c}$ of that element. 
$T_{c}$ express 
the volatility of the elements in dust \cite{LOD03}. 
In general, elements with higher condensation 
temperatures  are more likely to aggregate into dust grains and are more resistant to destruction. 
\cite{LOD03} has calculated the condensation temperatures for various elements for a solar-system composition gas. 
Her results indicate condensation temperatures of 78 K, 182 K and 704 K  for C, O and S, respectively. 
On the other hand, the condensation temperatures calculated for Fe, Si and Mg are 1357 K, 1529 and 1397, respectively, 
hence considerably higher than for C, O and S. \\
As shown by Fig.~\ref{fract_epsi}, by assuming for C, O and S a dust destruction efficiency $\epsilon_{C,O,S}=0.8$ and  for 
Fe, Si, and Mg  $\epsilon_{Fe,Mg,Si}=0.2$, it is possible to reproduce the observed dust fractions with good accuracy. \\

\begin{figure*}
\centering
\vspace{0.001cm}
\epsfig{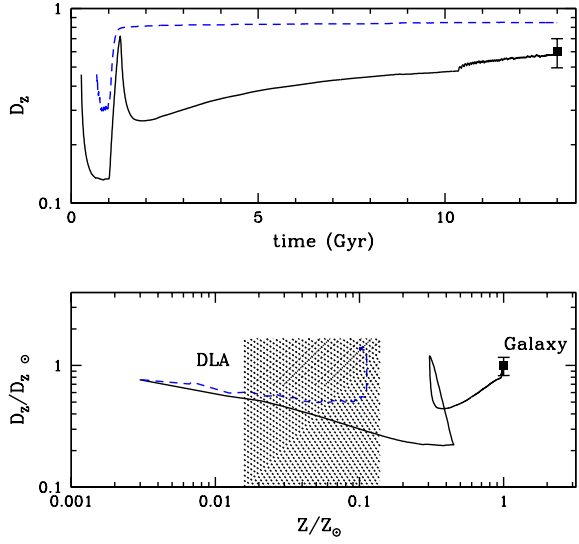}
\caption[]{Upper panel: predicted evolution of the dust-to-metal ratio in the S.N. (solid line) and 
at 16 Kpc from the Galactic centre (dashed line). The open square with error 
bar is the observational value for the Galaxy assuming the dust-to-gas ratio estimated by \cite{ISS90} and a solar metallicity 
of $Z_{\odot}=0.0133$ \cite{LOD03}.  
Lower panel: predicted dust-to-metal ratio as a function of the metallicity $Z$ in the S.N. (solid line) 
and at 16 kpc from the Galactic centre (dashed line). 
The dust-to-metal ratio and the metallicity are normalized to the Galactic value and 
to the solar 
value, respectively. The dotted region is the range observed in DLA systems \cite{VLA04}. 
The open square is the value observed in our Galaxy.}
\label{dtm}
\end{figure*}

In Figure ~\ref{dtm}, we show the time and metallicity evolution of the 
dust-to-metal ratio  $D_{Z}$ (defined as the ratio between the dust
mass and the metal mass surface density in the ISM) for the S.N. model and for for the outermost regions of the spiral disc. 
We compare our predictions to the values observed in the S.N. and in Damped Lyman Alpha (DLA) systems, 
which are believed to represent the high-redshift
counterparts of local gas-rich galaxies,  such as spirals and dwarf
irregulars \cite{CAL03a}. This figure shows that the model for the outskirts of a spiral disc 
presents dust-to-metal values and metallicities fully consistent with the values obsereved in DLAs.

\subsection{Dust evolution in elliptical galaxies}
\begin{figure*}
\centering
\vspace{0.001cm}
\epsfig{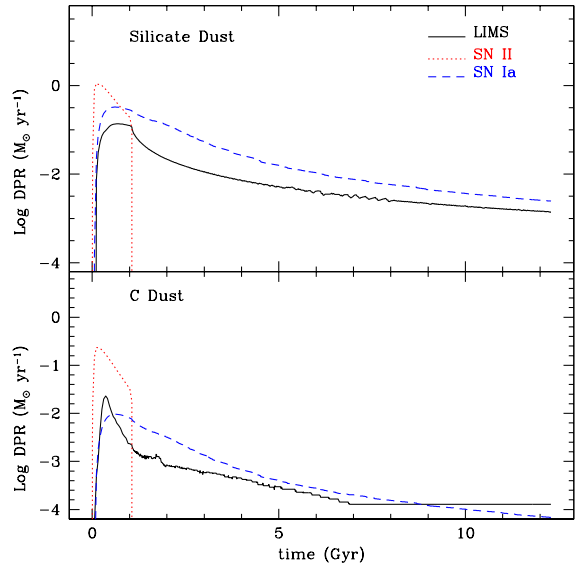}
\caption[]{Production rates for several dust sources for C and silicates predicted by model Ha1+Itoh.}
\label{dust_ell_4}
\end{figure*}

In this Section, we use the best values for the parameters derived above
by studying dust in the S.N., in order to extend our formalism to the class of elliptical galaxies. 
By looking at Fig.~\ref{dust_ell_4}, showing the predicted 
evolution of the dust production rates in an elliptical galaxy, we see that 
in ellipticals,  the  SNe Ia are the major dust factories in the last 10 Gyr.  \\

\begin{figure*}
\centering
\vspace{0.001cm}
\epsfig{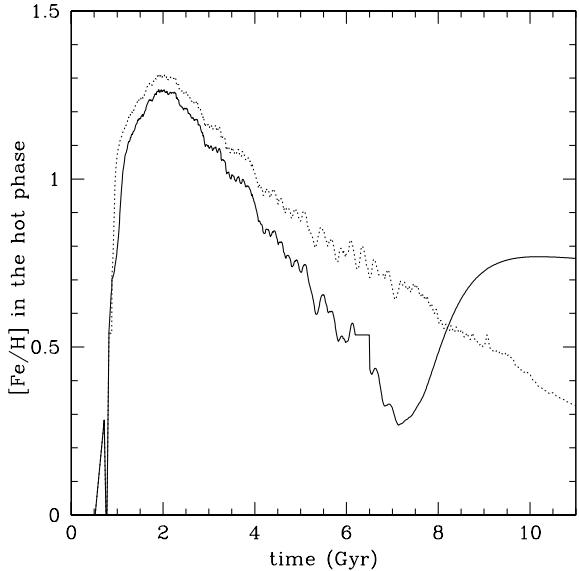} 
\caption[]{Evolution of the Fe abundance in the hot phase of an elliptical galaxy for the La1 model. 
Solid lines: La1+Itoh model; Dotted Line: La1+MK model.}
\label{dust_Fe_disc}
\end{figure*}

After the onset of the galactic wind, ellipticals are basically devoid of cold gas. Therefore,  
we assume that, during this phase, no dust accretion can occur in elliptical galaxies.  
We note that \cite{MCK89} estimates $\epsilon M_{SNR} \sim 70 M_{\odot}$ for a hot and rarefied medium as the gaseous halos
surrounding ellipticals, therefore we run two models in which $\epsilon M_{SNR} = 1360 M_{\odot}$ until the galactic wind
and then we have an instantaneous transition to $\epsilon M_{SNR} \sim 70 M_{\odot}$. The pre-wind
dust evolution is obviously unaffected, whereas it leads to substantial changes in the late
stage of the galactic evolution. According to the galaxy mass, we call them \emph{La1+MK} and \emph{Ha1+MK}, respectively. 
We further modified the destruction treatment by implementing a thermal sputtering term,
which is thought to be the dominant source of dust destruction in hot plasmas.
In particular, following \cite{ITO89} and \cite{ARI97},
we assume that, in a $\sim$1 keV plasma, nearly 90\% of the dust grains will evaporate
by thermal sputtering in $\tau_{destr_{sp}, i} \sim 10^5 /n_e (\rm yr\; cm^{-3})$, 
where the electron density $n_e$ has been self-consistently evaluated at each timestep.
This translates into a new destruction term, namely: 
\begin{equation}
\frac{G_{dust,i}}{\tau_{destr}}= G_{dust,i} (70 M_{\odot}) \frac{R_{SN}}{\sigma_{gas}} + \frac{G_{dust,i}}{\tau_{destr_sp,i}}
\end{equation} 
According to the galaxy mass, we call the models featuring this particular term as \emph{La1+Itoh} and \emph{Ha1+Itoh}, respectively.
When studying chemical evolution of the dust in ellipticals, it may be important to investigate whether the inclusion
of the dust may help in solving the so-called iron discrepancy \cite{ARI97}. 
The \emph{ASCA} satellite provided the first reliable measures of the iron
abundance in the hot ISM of ellipticals
(e.g. \cite{AWA94}; \cite{MAT97}), finding values much lower than solar, at odds not only with theoretical 
models for elliptical galaxies available at that time (\cite{ARI97}; \cite{MAT87}), which predicted that their ISM should exhibit
[Fe/H]$>0$, but also with the mean metallicity of the stellar component inferred from optical spectra. 
By considering the most recent data, the Fe discrepancy is partially alleviated, but 
clearly still present when comparing the 
chemical evolution results by \cite{PIP05}, 
predicting [Fe/H]$\ge 0.85 $ in the ISM of elliptical galaxies, to the abundance measurements in the X-ray spectra, 
indicating emission-weighted Fe abundances up to $\sim 2-3 $ solar in a sample of 28 early-type galaxies, corresponding to [Fe/H]=0.3 - 0.48 
\cite{HUM06}. \\
Fe condensation in dust is among the possible physical mechanisms often invoked to solve this issue (see
\cite{ARI97} for a comprehensive analysis).
Recent far-infrared observations, in fact, claim that the dust mass in ellipticals could be $\sim 10^{6-7}M_{\odot}$,
a factor of ten higher than previous estimates \cite{TEM04},
with $\sim 2\cdot 10^{5}M_{\odot}$ of dust residing in the very central galactic regions \cite{LEE04}.
Independent recent observations carried out by means of the SCUBA camera
on local ellipticals \cite{VLA05} seem to confirm these values, although the 
sample should be enlarged and the effects of synchrotron radiation clarified.
These new observational values are well reproduced by our models, which
predict a dust mass of $\sim 1.1 \cdot 10^{6}M_{\odot}$ for more than 10 Gyr old giant ellipticals  ($\sim 0.8 \cdot 10^{6}M_{\odot}$
of dust for a more typical galaxy like our model La1).\\
In Figure ~\ref{dust_Fe_disc}, we show the predicted  
evolution of the Fe abundance in the hot phase of a giant elliptical galaxy by adopting the \cite{ITO89} (solid line) and the 
\cite{MCK89} (dotted line) prescriptions. The best results have been obtained by adoptiong the Mckee presciptions for dust destruction. 
With these prescriptions, the predicted present-day iron abundance in the hot gas in the case of a $L_{*}$ and a giant elliptical is 
[Fe/H]=+0.3 and +0.15 dex, respectively. These values are in excellent agreement with the available observations. 

Moreover, since the Fe emission line dominates the X-ray spectrum of the hot ISM,
a reduced Fe abundance in the gas leads to a better agreement between the predicted X-ray luminosity
($2.1 \cdot 10^{42} \rm erg\, sec^{-1}$) and temperature ($\sim$ 0.7 kev) for the hot halo
of giant ellipticals with the observed values, than in \cite{PIP05} case.\\
To conclude, we suggest that the iron discrepancy might be solved 
by taking into account the dust condensation of Fe. New and elliptical-dedicated calculations
of the sputtering by hot (i.e. $T> 10^7$K) plasmas are needed to eventually asses this issue.

\subsection{Dust evolution in Irregular galaxies}

\begin{figure*}
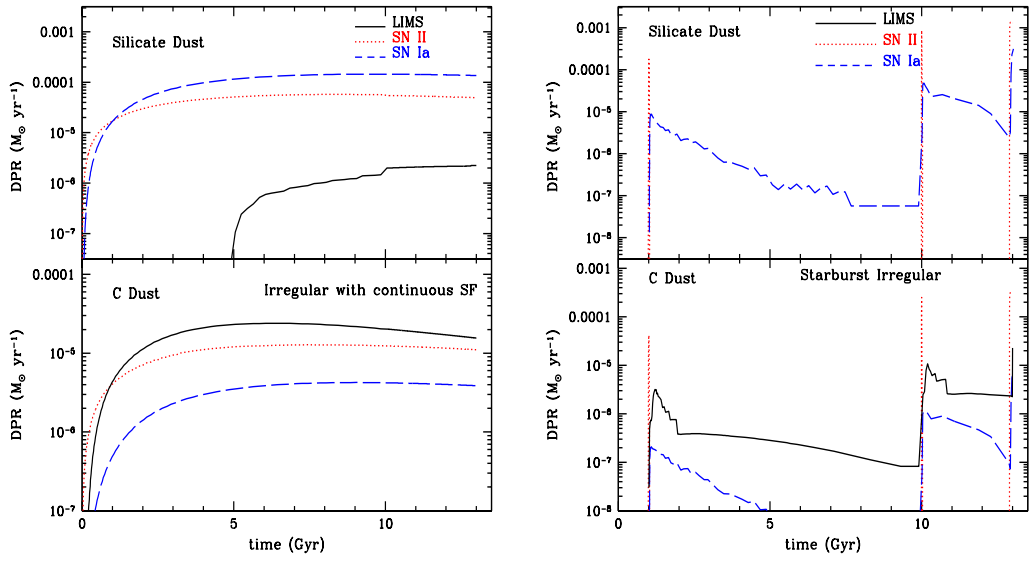

\centering
\vspace{0.001cm}
\leftline{\epsfig{file=fig7.eps,height=8cm,width=7cm}
\epsfig{file=fig8.eps,height=8cm,width=7cm} }
\caption[]{ Predicted dust production rates from various sources 
for a chemical evolution model of an irregular galaxy with continuous SF (left) and a starburst irregular galaxy (right).  
In the lower (upper) panels, we show the results for the carbon (silicate) dust. 
Solid lines: contribution by low and intermediate mass stars (LIMS). Dotted lines: contribution 
by type II SNe. Dashed lines: contribution by type Ia SNe.}
\label{IC+SB}
\end{figure*}

In Figure ~\ref{IC+SB}, 
we show the predicted dust production rates as functions of time for the models of an irregular galaxy with continuous SF (IC) 
and a starburst irregular.  

All the dust production rates calculated for the IC model have a smooth behaviour and in general present a continuous increase,  
lasting several Gyrs, and then reach a plateau. On the other hand, the production rates calculated for the starburst irregular 
reflect the shape of the SF and present a gasping or intermittent behaviour. 
An interesting difference between the dust production rates predicted for irregular galaxies and the one predicted for 
the S.N. model concerns the production of silicate dust by low and intermediate mass stars. 
In the case of the S.N. model, the Si dust production by LIMS starts immediately after the first SF episode. 
On the other hand, for the IC model Si dust production by LIMS starts at $\sim 5 $ Gyr and for the starburst model it does 
not start at all. This depends on the adopted prescriptions for the dust condensation efficiencies and on the C/O ratio in dwarf irregulars, 
as  can be seen in \cite{CAL08}.

\begin{figure*}
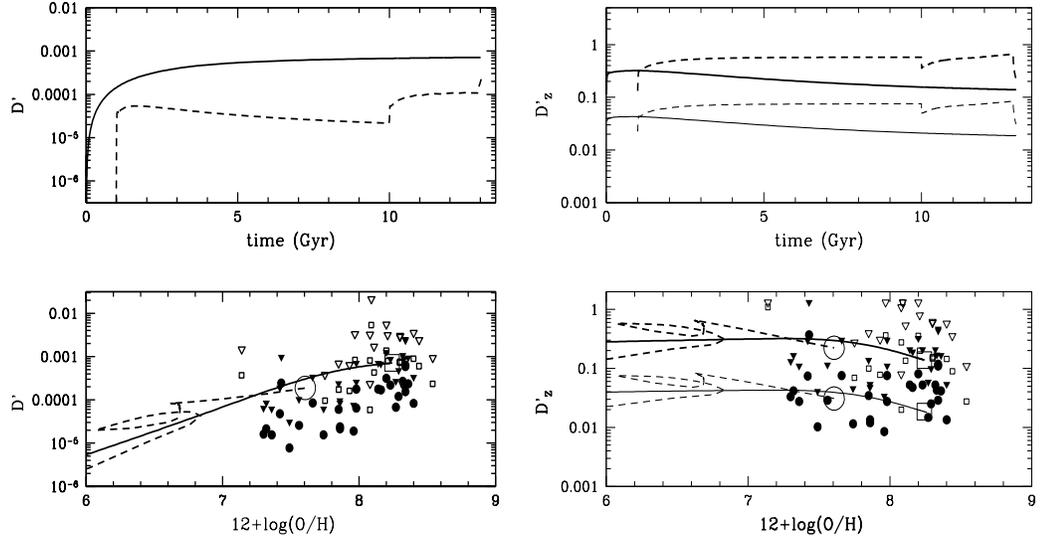

\leftline{\epsfig{file=fig9.eps,height=8cm,width=7cm}
\epsfig{file=fig10.eps,height=8cm,width=7cm} }
\caption[]{ \emph{Left:} Evolution of the dust-to-gas ratio calculated for the IC model (solid lines) 
and for the starburst model (dashed lines) as a function of time (upper panel) and metallicity (lower panel). 
In the lower panel, the solid circles and open squares are 
$D'$ values observed in local dIrr and BCD galaxies, respectively \cite{LIS98}. 
The solid and empty inverted triangles represent upper limits 
to the observations for local dIrr and BCD galaxies, respectively, calculated by assuming an uncertainty of a factor of 4.  
\emph{Right:} Evolution of the dust-to-metal ratio calculated for the IC model 
(solid lines) and for the starburst model (dashed lines) as a function of time (upper panel) and metallicity (lower panel). 
The thick and thin lines represent the predicted $D'$ calculated assuming for the condensation efficiencies 
the values  suggested by \cite{DWE98} and   
a constant value of 0.1, respectively. 
All the symbols are as in the left lower panel. }
\label{dtg_dtm_irr}
\end{figure*}

When studying dust evolution in irregular galaxies, 
it may be interesting to study the dust-to-gas and dust-to-metal ratios in order 
to compare the predictions with some directly-measurable quantities.
The dust-to-gas ratio is defined as: 
\begin{equation}
D'= \frac{M_{dust}}{M_{HI}}
\end{equation}
where $M_{dust}$ and $M_{HI}$ are the dust and $HI$ mass determined from the far infrared  
and from the radio emission, respectively \cite{LIS98}. 
In the models, the neutral H mass is $M_{HI}=M_{gas} \cdot X_{H}$, where $M_{gas}$ is the total gas mass and 
$X_{H}$ is the H mass fraction. 
In Figure~\ref{dtg_dtm_irr}, we show the $D'$  calculated for the IC model and for the starburst model as 
a function of time (upper left panel) and as a function of the metallicity (lower left panel), 
and compared to a set of observations in local dwarf galaxies. 
At any time, the IC model is characterised by higher $D'$ values than the starburst model. \\
In the lower panels of Figure~\ref{dtg_dtm_irr}, 
as a proxy of the metallicity we use the quantity 12+log(O/H). 
The observational values used here are taken from a compilation by \cite{LIS98} (see caption of Fig. ~\ref{dtg_dtm_irr}), 
and refer to BCD (solid circles) and dIrr (open squares) galaxies.  
We note that the predicted present-day values are in general higher than the majority of the observed values. 
The compilation of observations presented by \cite{LIS98} is based on IRAS data. 
The observed determinations of the dust mass are 
likely to represent lower limits to the actual values, owing to the undetected cold dust component 
(\cite{DWE98}, \cite{POP02}). 
\cite{LIS98} estimate a total error in the dust-to-gas ratio to be of a factor of 4, taking into account all the possible 
factors of uncertainty (contribution from very small grains, cold dust, molecular gas and variations in the $HI$ and optical  diameter). 
By taking into account this uncertainty (solid and open inverted triangles in Fig.~\ref{dtg_dtm_irr}), 
the observed dust-to-gas ratios are consistent with our predictions for  both irregular types. 

The dust-to-metals ratio $D'_{Z}$ observed in 
dwarf irregulars can be defined as: 

\begin{equation}
D'_{Z} = \frac{M_{dust}}{M_{HI} \cdot Z} = \frac{D'}{Z}
\end{equation} 

In dwarf irregular galaxies, dust accretion is assumed to be negligible, hence the dust fractions are determined from the balance 
between the dust condensation efficiencies and the dust destruction process. 
In the right upper and lower panels of Figure~\ref{dtg_dtm_irr}, we show the predicted evolution of $D'_{Z}$ in dwarf irregular and starburst galaxies
as a function of time and metallicity, respectively, 
calculated assuming the set of 
condensation efficiencies as suggested by \cite{DWE98} (thick lines) and a constant value 
of $\delta^{SW}_{i}=\delta^{Ia}_{i}=\delta^{II}_{i}=0.1$ (thin lines). 
The adoption of different condensation efficiencies has noticeable effects 
on the dust-to-metal ratio. In particular, the  $D'_{Z}$  values calculated by assuming the condensation efficiencies by \cite{DWE98} 
are larger than the ones predicted by assuming a constant value of 0.1.  \\

In the lower right panel of Fig.~\ref{dtg_dtm_irr}, we show the predicted evolution of $D'_{Z}$ as a function 
of the metallicity for the IC model (solid lines) and for the starburst model (dashed lines). 
The big open circles and the big open squares are the predicted present-day values for $D'_{Z}$ 
for the starburst and IC model, respectively. 
In the S.N., the dust fractions are determined by the balance between dust destruction and dust accretion, hence they are independent 
on the dust condensation efficiencies. Fig.~\ref{dtg_dtm_irr} shows that, 
in principle, 
the measure of the dust-to-metal ratio or of the dust depletion pattern in dwarf irregulars could be helpful 
to put solid constraints on the dust condensation efficiencies.\\

\section{Summary and future perspectives}
By means of chemical evolution models for galaxies of different morphological types, 
we have performed a study of the cosmic evolution of the dust properties in different environments. 
By starting from the same formalism as developed 
by \cite{DWE98}, we have carried on a deep study of the space of the parameters used to model 
dust evolution and, thanks to the uptodate observations available in the solar vicinity, 
performed a fine tuning of the parameters.  
We have extended our study to ellipticals and dwarf irregular galaxies for which, for the first time, 
dust evolution has been calculated by means of chemical evolution models relaxing the instantaneous recycling 
approximation, namely taking into account the stellar lifetimes. 
Our main results can be summarised as follows. \\
1) The evolution of the dust production rates and of the dust fractions are strongly determined by the SF history. \\
2) We can reproduce the dust fractions observed in the solar neighbourhood by assuming that 
the dust destruction efficiency depends the condensation temperatures $T_{c}$ of the elements. 
In general, more volatile elements have lower $T_{c}$ and higher dust destruction efficiencies. \\
3) In elliptical galaxies, type Ia SNe are the major dust factories in the last 10 Gyr. 
With our models, we successfully reproduce the dust masses observed in local ellipticals ($\sim 10^{6}M_{\odot}$)
by means of recent FIR and SCUBA observations. \\
4) We have shown that, by considering a reduced dust destruction in 
a hot and rarefied medium, dust is helpful in solving the iron discrepancy observed in the 
hot gaseous halos surrounding local ellipticals \cite{ARI97}. In this medium, we predict a
Fe abundance of [Fe/H]=0.15-0.3, whereas without dust \cite{PIP05} found [Fe/H]$\ge 0.85$.  \\
5) The two models used to study dust evolution in dwarf irregular galaxies
present very different dust production rates. For the IC model, the production rates have a smooth behaviour 
throughout the whole evolution of the system. For the starburst model, the evolution of the 
production rates reflects its intermittent SF history. \\
6) The predicted present-day dust-to-gas ratios for irregular galaxies  
are in general higher than the majority of the observed values. 
However, the determination of the dust mass in these galaxies is affected by several sources of uncertainty.  
By assuming a factor of 4 uncertainty in the observed data, as suggested by \cite{LIS98}, 
the observed dust-to-gas ratios are consistent with our predictions. \\
7) In dwarf irregulars, dust accretion is likely to play a negligible role or to be absent, hence the variation of the condensation 
efficiencies has important effects on the dust-to-metal ratio. 
In principle, a precise determination of the dust-to-metal ratios or of  the dust depletion pattern  
in dwarf irregular galaxies could be helpful to put solid constraints on the dust condensation efficiencies. \\
In the future, our results  
will be very useful for the study of the spectral properties of dust grains in galaxies of different morphological types. 
By combining the chemical evolution results described here with a spectrophotometric code that includes 
reprocessing by graphite and silicate dust grains (GRASIL, \cite{SIL98}), it will be possible 
to analyse the spectral energy distributions of galaxies of different morphological types (Schurer et al. 2008, in prep.) and to 
considerably improve our understanding of interstellar dust.


\begin{thebibliography}{99}
\bibitem{ARI97} Arimoto N., Matsushita K., Ishimaru Y., Ohashi T.,  Renzini,  A. 1997, ApJ, 477, 128
\bibitem{AWA94} Awaki H., Mushotzky R., Tsuru T., Fabian  A.C., Fukazawa Y., Loewenstein M., Makishima  K., Matsumoto H. et al. 1994, PASJ, 46, L65
\bibitem{BRA98} Bradamante F., Matteucci F., D'Ercole A., 1998, A\&A, 337, 338    	
\bibitem{CAL03} Calura, F., Matteucci, F., 2003,  ApJ, 596, 734
\bibitem{CAL03a} Calura F., Matteucci F., Vladilo G., 2003, MNRAS, 340, 59 		 
\bibitem{CAL04} Calura, F., Matteucci, F., 2004, MNRAS, 350, 351
\bibitem{CAL04a} Calura, F., Matteucci, F., Menci, N., 2004, MNRAS, 353, 500
\bibitem{CAL06a} Calura F., Matteucci F., 2006a, MNRAS, 369, 465
\bibitem{CAL06b} Calura F., Matteucci F., 2006b, ApJ, 652, 889
\bibitem{CAL08} Calura, F., Pipino, A., Matteucci, F., 2008, A\&A, in press,  arXiv0706.2197C
\bibitem{CHI97} Chiappini, C., Matteucci, F., Gratton, R. 1997, ApJ, 477, 765 
\bibitem{CHI01} Chiappini, C., Matteucci, F., Romano, D., 2001, ApJ, 554, 1044
\bibitem{CLA96} Clayton, Geoffrey C., Green, J., Wolff, Michael J., Zellner, Nicolle E. B., Code, A. D., Davidsen, Arthur F., WUPPE Science Team, HUT Science Team, 1996, ApJ, 460 313 
\bibitem{DRA03} Draine B. T., 2003, ARA\&A, 41, 241
\bibitem{DWE98} Dwek E., 1998, ApJ, 501, 643 
\bibitem{FRA04} Fran\c cois P., Matteucci F., Cayrel R., Spite M., Spite F., Chiappini C., 2004, A\&A, 421, 613
\bibitem{GRE83} Greggio L., Renzini A., 1983, A\&A, 118, 217     	
\bibitem{HUM06} Humphrey P. J., Buote D. A., 2006, ApJ, 639, 136
\bibitem{INO03} Inoue A. K., 2003, PASJ, 55, 901
\bibitem{ISS90} Issa M. R., MacLaren I., Wolfendale A. W., 1990, A\&A, 236, 237
\bibitem{ITO89} Itoh H., 1989, PASJ, 41, 853
\bibitem{IWA99} Iwamoto K., Brachwitz F., Nomoto K., Kishimoto N., Umeda H., Hix W. R., Thielemann, F.-K., 1999, ApJS, 125, 439I
\bibitem{KIM03} Kimura H., Mann I., Jessberger E. K., 2003, ApJ, 582, 846
\bibitem{LEE04} Leeuw L. L., Sansom A. E., Robson E. I., Haas M., Kuno N., 2004, ApJ, 612, 837
\bibitem{LIS98} Lisenfeld U., Ferrara A., 1998, ApJ, 496, 145
\bibitem{LOD03} Lodders K., 2003, ApJ, 591, 1220
\bibitem{MAT97} Matsumoto H., Koyama K., Awaki H., Tsuru T., Lowenstein M., Matsushita K., 1997, ApJ, 482, 133
\bibitem{MAT94} Matteucci F., 1994, A\&A, 288, 57                     
\bibitem{MAT86} Matteucci F., Greggio L., 1986, A\&A, 154, 279
\bibitem{MAT87} Matteucci F., Tornamb\'e A., 1987, A\&A, 185, 51 
\bibitem{MAT89} Matteucci F., Fran\c cois P., 1989, MNRAS, 239, 885 
\bibitem{MCK89} McKee C. F., 1989, in Allamandola L. J., Tielens A. G. G. M., eds, Interstellar Dust, Proc. IAU Symposium 135. Kluwer, Dordrecht, p. 431
\bibitem{PIP02} Pipino A., Matteucci F., Borgani S., Biviano A., 2002, NewA, 7, 227	
\bibitem{PIP05} Pipino A., Kawata D., Gibson B. K., Matteucci F., 2005, A\&A, 434, 553	
\bibitem{POP02} Popescu C. C., Tuffs R. J., Voelk H. J., Pierini D., Madore B. F., 2002, ApJ, 567, 221
\bibitem{SAL55} Salpeter E.E., 1955, ApJ, 121, 161
\bibitem{SCA86} Scalo J. M., 1986, FCPh, 11, 1
\bibitem{SIL98} Silva, L., et al., 1998, ApJ, 509, 103
\bibitem{TEM04} Temi P., Brighenti F., Mathews W. G., Bregman J. D., 2004, ApJS, 151, 237
\bibitem{THO98} Thornton K., Gaudlitz  M., Janka  H.-T.,  Steinmetz M. 1998, ApJ, 500, 95
\bibitem{VAN97} van den Hoek L. B. \& Groenwegen M. A. T., 1997, A\&AS, 123, 305
\bibitem{VLA05} Vlahakis C., Dunne L., Eales S., 2005, MNRAS, 364, 1253
\bibitem{VLA04} Vladilo G., 2004, A\&A, 421, 479
\bibitem{WOO95} Woosley S.E., Weaver T.A., 1995, ApJS, 101, 181
\bibitem{ZUB04} Zubko V., Dwek E., Arendt R. G., 2004, ApJS, 152, 211
\end{thebibliography}
\end{document}